\documentclass[reprint, aps,prl,twocolumn,superscriptaddress]{revtex4-1}
\pdfoutput=1
\usepackage{amssymb}
\usepackage{amsmath}
\usepackage{epsfig}
\usepackage{color}
\usepackage{graphics, graphicx}
\usepackage{bbold}
\usepackage{psfrag}
\usepackage{mathcomp}
\usepackage{subfigure}
\usepackage{verbatim}
\usepackage[colorlinks, citecolor=blue]{hyperref}
\usepackage[normalem]{ulem}
\def\cp#1{\mathbf{#1}}
\begin{document}

\title{Borromean droplet in three-component ultracold Bose gases}
\author{Yinfeng Ma}
\affiliation{Beijing National Laboratory for Condensed Matter Physics, Institute of Physics, Chinese Academy of Sciences, Beijing 100190, China}
\affiliation{School of Physical Sciences, University of Chinese Academy of Sciences, Beijing 100049, China}
\author{Cheng Peng}
\affiliation{Beijing National Laboratory for Condensed Matter Physics, Institute of Physics, Chinese Academy of Sciences, Beijing 100190, China}
\affiliation{School of Physical Sciences, University of Chinese Academy of Sciences, Beijing 100049, China}
\author{Xiaoling Cui}
\email{xlcui@iphy.ac.cn}
\affiliation{Beijing National Laboratory for Condensed Matter Physics, Institute of Physics, Chinese Academy of Sciences, Beijing 100190, China}
\affiliation{Songshan Lake Materials Laboratory, Dongguan, Guangdong 523808, China}
\date{\today}

\begin{abstract}
We investigate the droplet formation in three-component ultracold bosons. In particular, we identify the formation of {\it Borromean droplet}, where only the ternary bosons can form a self-bound droplet while any binary subsystems cannot, as the first example of Borromean binding due to collective many-body effect.  Its formation is facilitated by an additional attractive force induced by the density fluctuation of a third component, which enlarges the mean-field collapse region in comparison to the binary case and renders the formation of Borromean droplet after incorporating the repulsive force from quantum fluctuations. Outside the Borromean regime, we demonstrate an interesting phenomenon of droplet phase separation due to the competition between ternary and binary droplets. We further show that the transition between different droplets and gas phase can be conveniently tuned by boson numbers and interaction strengths. The study reveals rich physics of quantum droplet in three-component boson mixtures and sheds light on more intriguing many-body bound state formed in multi-component systems.
\end{abstract}
\maketitle

{\it Introduction.}
Discovering peculiar bound states helps to expand our horizons in understanding intriguing quantum  effect in a physical world. The Borromean binding clearly belongs to such case, where only three items together can form the bound state while any two of them cannot. It has been successfully reported in nuclear physics as halo nuclei\cite{halo1,halo2} and in ultracold gases as the Efimov effect\cite{Efimov_review1,Efimov_review2,Efimov_review3}. In these occasions, the Borromean binding refers to the trimer formation in few-body clusters where no dimer is present, such as the Efimov trimers observed in the negative scattering length side\cite{Efimov_Exp0,Efimov_Exp1,Efimov_Exp2,Efimov_Exp3,Efimov_Exp4,Efimov_Exp5,Efimov_Exp9,Efimov_Exp10} that are supported by the attractive potential due to quantum interference of three particles. Theoretical studies have found that the Borromean trimer can be equally supported by fine-tuning the shape and strength of pairwise potential\cite{Richard, Moszkowski,Nielsen, Volosniev,Volosniev2} or by modifying the single-particle dispersion\cite{Cui}.  Given the stringent requirement for its occurrence in small clusters, whether the Borromean binding can be extended to many-body systems due to collective effect is an interesting yet challenging problem.  

On the other hand, as a typical many-body bound state, droplet has been well studied in helium liquid\cite{helium_expt,helium_theory} and in Bose-Einstein systems with short- and long-range interactions\cite{Huang}. Recently, it has regained great attention in ultracold atoms following a pioneer proposal by Petrov\cite{Petrov}. Stabilized by the mean-field attraction and the Lee-Huang-Yang(LHY) repulsion from quantum fluctuations, quantum droplet has so far been successfully observed in dipolar gases\cite{Pfau_1,Pfau_2,Pfau_3,Ferlaino,Modugno,Pfau_4,Ferlaino_2} and binary Bose gases of alkali atoms\cite{Tarruell_1,Tarruell_2,Inguscio,Modugno_2}. It has also been theoretically extended to low dimensions\cite{Petrov_2, Santos, Jachymski, Zin, Buchler, Cui3}, Bose-Fermi mixtures\cite{Cui2, Adhikari, Rakshit1, Rakshit2, Wenzel,Yi}, dipolar mixtures\cite{Blakie,Santos_2} etc.

In this work, we point out the first example of Borromean binding due to collective many-body effect,  namely, the {\it Borromean droplet} in three-component boson mixtures. Specifically, ``Borromean" means that only ternary  bosons can form the droplet while any binary subsystems cannot. Its physical origin lies in an additional attractive force induced by the density fluctuation of a third component, which further intensifies the mean-field collapse and renders the formation of Borromean droplet after incorporating the LHY repulsive force.  Such collective effect is substantially different from the mechanism of Borromean binding in small clusters\cite{Efimov_review1,Efimov_review2,Efimov_review3, Richard, Moszkowski,Nielsen, Volosniev,Volosniev2, Cui}.
Outside the Borromean regime, we demonstrate an interesting phenomenon of droplet phase separation due to the competition between ternary and binary droplets. 
The emergence of these different droplets, which is shown to be conveniently tuned by the species number and the coupling strengths, shed light on more intriguing bound state formation in multi-component systems. 

{\it Model.} We start with the Hamiltonian for three-component bosons
$H=\int d{\bf r} H({\bf r})$, with ($\hbar=1$)
\begin{equation}
H({\bf r})=\sum_{i=1,2,3}\Psi_i^{\dag}({\bf r})(-\frac{\nabla^2}{2m_i})\Psi_i({\bf r})+\sum_{ij}\frac{g_{ij}}{2} \Psi_i^{\dag}\Psi_j^{\dag}\Psi_j\Psi_i({\bf r}).
\end{equation}
Here ${\bf r}$ is the coordinate; $m_i$ and $\Psi_i$ are respectively the mass and field operator of boson species $i$;  
$g_{ij}$ is the s-wave coupling strength between species $i$ and $j$.

For a homogeneous system with uniform densities $\{n_i\} \ (i=1,2,3)$, the mean-field energy per volume is given by
\begin{equation}
\epsilon_{\rm mf}=\frac{1}{2}\sum_{i,j=1}^3 g_{ij}n_in_j
\end{equation}
Following the standard Bogoliubov theory to treat quantum fluctuations\cite{book}, we obtain the LHY energy per volume as:
\begin{equation}
\epsilon_{\rm LHY}=\int\frac{d^3{\cp k}}{2(2\pi)^3}\left[\sum_{i=1}^3 (E_{i{\cp k}}-\epsilon_{i{\cp k}}-g_{ii}n_i)+\sum_{ij} \frac{2m_{ij} g_{ij}^2n_in_j}{{\cp k}^2}\right]. \label{e_LHY}
\end{equation}
Here $E_{i{\cp k}}$($i=1,2,3$) are the Bogoliubov spectra\cite{supple}.

To describe a droplet with inhomogeneous densities, we  adopt the local density approximation(LDA) and write the total LHY energy as $E_{\rm LHY}=\int d{\cp r} \epsilon_{\rm LHY}(n_i({\cp r}))$, with $n_i({\cp r})=|\Psi_i({\cp r})|^2$. This leads to three coupled Gross-Pitaevskii(GP) equations for $\{\Psi_{i}({\cp r})\}$ ($i=1,2,3$):
\begin{eqnarray}
i\partial_t\Psi_i&=&\left[-\frac{\nabla^2}{2m_i}+\sum_j g_{ij}|\Psi_j|^2+\frac{\partial \epsilon_{\rm LHY}}{\partial n_i}\right]\Psi_i, \label{GP}
\end{eqnarray}
The ground state can be approached by the imaginary time evolution of above equations.

In this work, to facilitate discussions while keeping the essence of physics, we consider the equal mass case  $m_i\equiv m$ and the coupling strengths with following symmetry
\begin{equation}
g_{11}=g_{22}\equiv g, \ \ \ \ \ g_{13}=g_{23}\equiv g'. \label{simp}
\end{equation}

{\it Mean-field stability.} We first analyze the mean-field stability against density fluctuations of three-component(ternary) bosons, and compare it with the two-component(binary) cases. 
The stability is determined by the second-order variation of $\epsilon_{\rm mf}$ with respect to small change of local densities $\delta n_i$: $\delta^2 \epsilon_{\rm mf}=\sum_{ij} \frac{1}{2}g_{ij}\delta n_i\delta n_j$, which gives:
\begin{equation}
\delta^2 \epsilon_{\rm mf}=\frac{g-g_{12}}{2}\delta n_-^2+\frac{g+g_{12}}{2}\delta n_+^2+\frac{g_{33}}{2}\delta n_3^2+\sqrt{2} g'\delta n_3\delta n_+, \label{mf_fluc}
\end{equation}
with $\delta n_{\pm}\equiv (\delta n_1\pm\delta n_2)/\sqrt{2}$ the {\it diagonalized} fluctuation modes for components 1 and 2. Eq.\ref{mf_fluc} clearly shows that the fluctuation of component 3 will interfere with  $\delta n_{+}$ and result in two new eigen-modes, while $\delta n_{-}$ is left unchanged. The mean-field stability requires $\delta^2 \epsilon_{\rm mf}>0$ for any $\delta n_i$, which leads to the following condition for a stable ternary system:
\begin{equation}
g>|g_{12}|;\ \ \ g_{33}>\frac{2g'^2}{g+g_{12}}. \label{stable_3}
\end{equation}
Note that the first condition ensures the stability of ($1\&2$) system, while the second one is due to the interference between (3) and ($1\&2$) and ensures the stability of ($1\&2\&3$). Importantly, compared to the stability condition $g'^2<gg_{33}$ for ($2\&3$) or ($1\&3$), the requirement in (\ref{stable_3}) is more stringent. Therefore, there exists a finite parameter window
\begin{equation}
\frac{g'^2}{g_{33}}\in(\frac{g+g_{12}}{2},g),\label{Borro_cond}
\end{equation}
such that all binary subsystems are stable against density fluctuations while ($1\&2\&3$) is not.

The intensified mean-field instability of ternary bosons, as compared to all binary subsystems, can be attributed to the additional attractive force brought by the third component. To see this efficiently, let us consider a special case with $g_{12}=0$ and start from the subsystem ($1\&3$) whose stability condition is $g'^2<gg_{33}$. This condition can be re-formulated as $g_{33}+g_{ind}>0$, with $g_{ind}=-g'^2/g$ the induced interaction (to component 3) by the density fluctuation of component 1\cite{book}. Now if add the component 2 to ($1\&3$), the fluctuation of component 2 will induce an additional attraction to component 3 and now $g_{ind}=-2g'^2/g$ is doubled. Thus $g_{33}$ needs to be more repulsive than in ($1\&3$) case  in order to stabilize ($1\&2\&3$). For a finite $g_{12}$, the fluctuations of 1 and 2 will couple together and give $g_{\rm ind}=-2g'^2/(g+g_{12})$, again more attractive than the binary cases.  We have checked that such enhanced $g_{\rm ind}$ in ternary system robustly applies for more general coupling strengths beyond (\ref{simp}).

\begin{figure}[t]
\includegraphics[width=8cm]{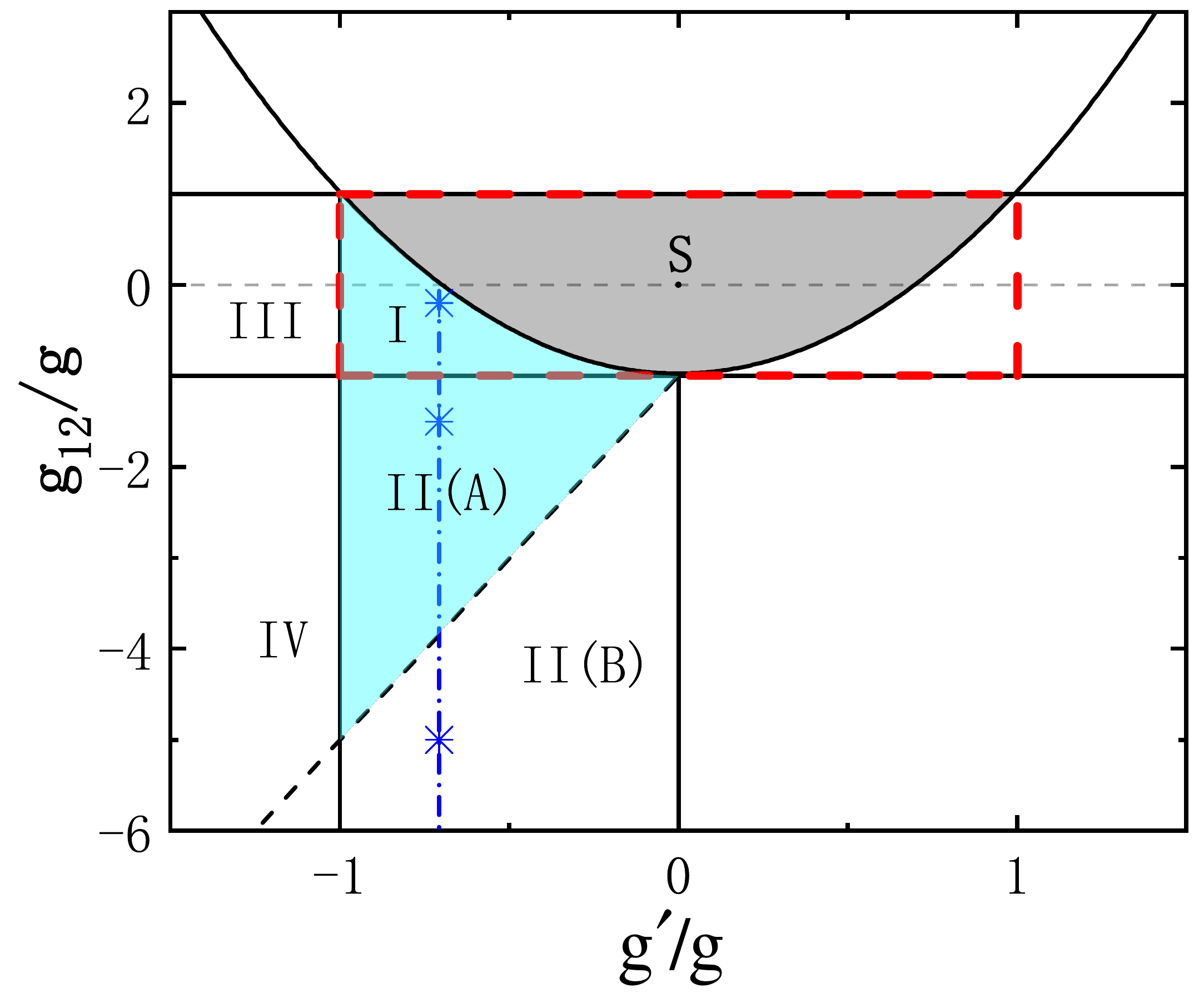}
\caption{(Color online) Mean-field phase diagram for three-component bosons under couplings (\ref{simp}) and  $g_{33}=g$. The gray area marks the mean-field stable region(``S"), smaller than that for binary subsystems (bounded by red square). 
After incorporating the LHY repulsion, Borromean droplet can take place in regions I and II(A). The dashed lines separating II(A) and II(B) is determined by $C_{\rm min}=0$ at the droplet-gas transition (see text). 
} \label{fig_mf}
\end{figure}

In Fig.\ref{fig_mf}, we plot out the mean-field phase diagram of ($1\&2\&3$) system in ($g',g_{12}$) plane taking a fixed $g_{33}=g$. The mean-field stable region, as required by (\ref{stable_3}) and labeled as ``S", is shown to be smaller than the stable region of binary subsystems (bounded by red square). For other regions in the diagram, a homogeneous $(1\&2\&3)$ system will undergo a collapse or phase separation due to density fluctuations, as determined by the eigen-modes of (\ref{mf_fluc})\cite{supple}. Among them, there are four regions, labeled as I,II,III,IV in Fig.\ref{fig_mf}, that all the three components undergo collapse simultaneously. This offers a possibility for droplet formation when further incorporating the repulsive force from quantum fluctuations. Of particularly interesting is regions I and II, as analyzed below.

{\it Borromean droplet.}  It is obvious that in region I, which satisfies  (\ref{Borro_cond}), the ternary system ($1\&2\&3$) can form a self-bound droplet while any binary subsystem cannot. By definition, this is the {\it Borromean droplet}. Meanwhile, since the actual droplet formation also depends on particle numbers $\{N_i\}$, it is possible for such intriguing state existing in other regions by properly tuning $\{N_i\}$. Given the coupling symmetry in Eq.(\ref{simp}), we have  $N_1=N_2$ for the ground state and there left two tunable parameters  for  $\{N_i\}$: total number $N=2N_1+N_3$ and number ratio $C=N_3/N_1$.

\begin{figure}[t]
\includegraphics[width=9cm]{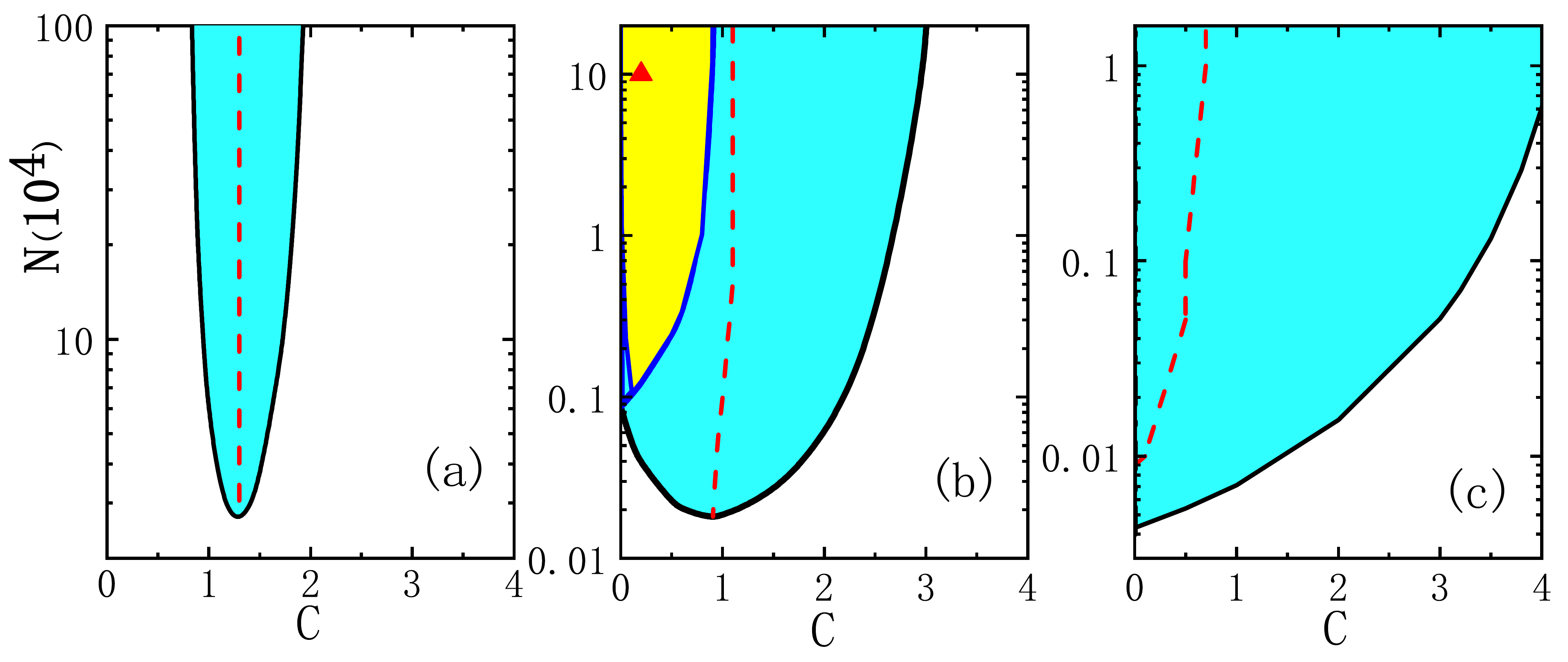}
\caption{(Color online.) Droplet region(colored) in the $N$-$C$ plane for three typical points (marked by ``*") on the vertical line in Fig.\ref{fig_mf}, which have a fixed $g'/g=-\sqrt{2}/2$ and  different $g_{12}/g=-0.2(a);  -1.5(b); -5(c)$. 
Red dashed lines show $C_{\rm min}$ when the system has minimal total energy at given $N$. 
In (b), the yellow area marks the regime for droplet phase separation. } \label{fig_Borro}
\end{figure}

\begin{figure}[t]
\includegraphics[width=7cm]{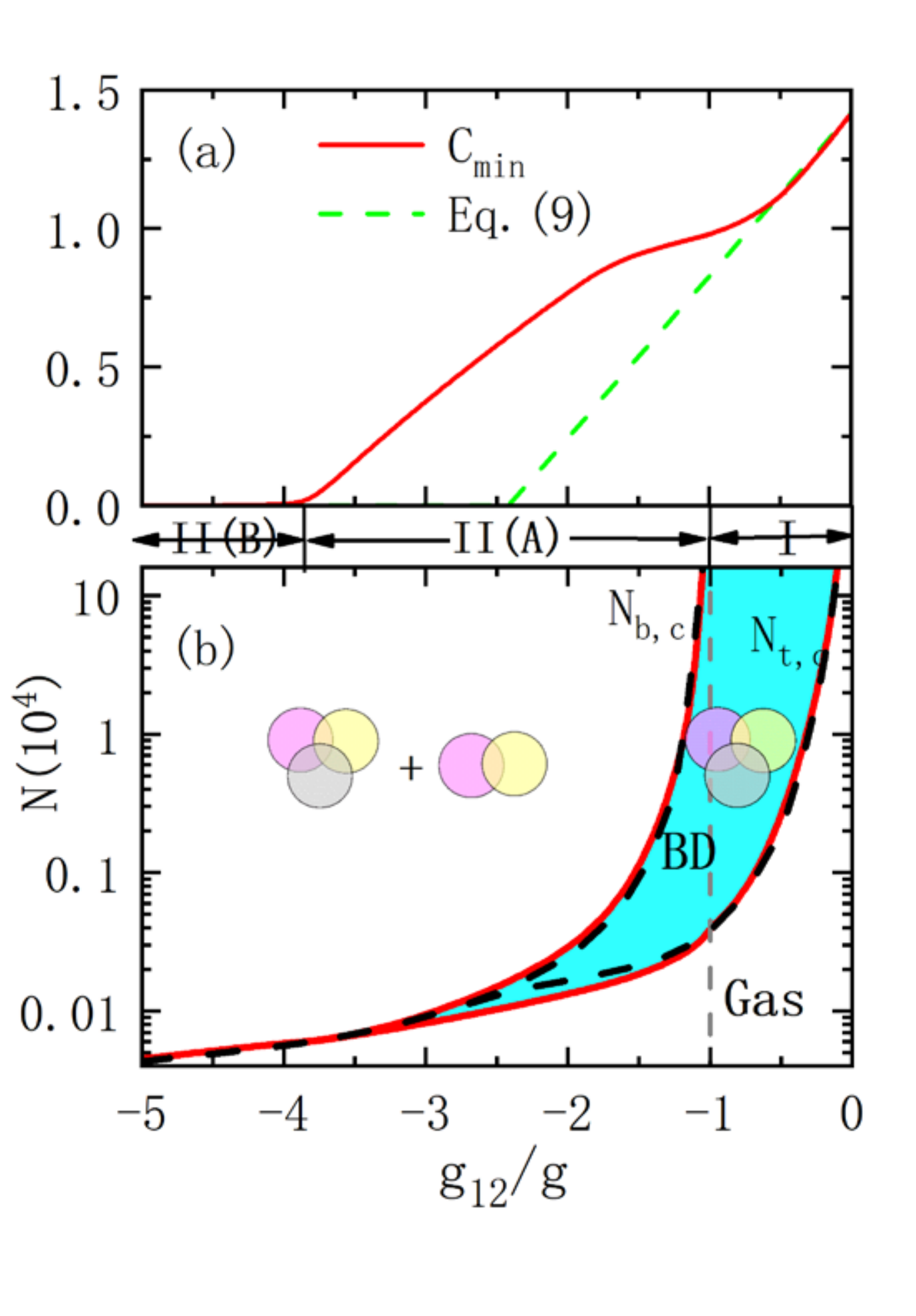}
\caption{(Color online.) (a)$C_{\rm min}$(solid line) as a function of $g_{12}/g$ at the critical number $N_{t,c}$, in comparison with $C_{\rm min}^{(0)}$(dashed line) from Eq.(\ref{c_eq}). (b) Phase diagram in the $N$-$g_{12}$ plane. 
The Borromean droplet (``BD") occurs in region I for $N>N_{t,c}$ and in II(A) for $N\in(N_{t,c},N_{b,c})$. For $N<N_{t,c}$ the system is in gas phase; for $N>N_{b,c}$ the binary droplet can also exist. Dashed lines show the function fits of $N_{t,c}^{(0)}$ (Eq.\ref{N3c}) and $N_{b,c}^{(0)}$ (from Ref.\cite{Petrov}).  Here $g'/g=-\sqrt{2}/2$.} \label{fig_phase}
\end{figure}

To explore essential properties of Borromean droplet, we carry out full simulations of the GP equations (\ref{GP}) to search for ground state with a fixed $g'/g=-\sqrt{2}/2$ and different $g_{12}/g$ in regions I and II, i.e., following the vertical line in Fig.\ref{fig_mf}.
In Fig.\ref{fig_Borro}(a-c), we show the area of droplet formation in the $N$-$C$ plane for three typical values of $g_{12}/g$, where we also show the value of $C$ when the energy reaches minimum for each $N$, denoted as $C_{\rm min}$ (dashed lines in Fig.\ref{fig_Borro}). One can see that for $g_{12}/g$ in region I (Fig.\ref{fig_Borro}(a)), a ternary droplet can be supported when $N$ is beyond a critical number, $N_{t,c}$, where a gas to droplet transition occurs. For all $N>N_{t,c}$, the droplet can only survive for $C$ within a narrow window around $C_{\rm min}$. This characterizes the Borromean nature of the droplet, i.e., its formation cannot extend to $C=0$ (when the third component is absent).

Interestingly,  the Borromean droplet can extend to part of the region II. As an example, for the parameters considered in Fig.\ref{fig_Borro}(b), we can see that as increasing $N$, a ternary droplet  first emerges at $N_{t,c}$ with a finite $C$. As $N$ is further increased to $N_{b,c}$,  the binary ($1\&2$) droplet appears at $C=0$. 
It is then followed that the Borromean droplet is stabilized within the number window $N\in(N_{t,c},N_{b,c})$, where $N$ is large enough to support a ternary droplet but still small for the binary one. 
The Borromean droplet will vanish when go deep into region II. As shown in Fig.\ref{fig_Borro}(c) for large attractive $g_{12}$, as increasing $N$ the droplet solution first emerges at $N_{b,c}$ with $C=0$. In this case, a binary droplet is more favored than a ternary one. 

Given the different behaviors of droplet formation in region II, we have separated this region into II(A) and II(B) in Fig.\ref{fig_mf} --- the former can support the Borromean droplet (within certain number window) while the latter cannot. Their boundary (dashed line in Fig.\ref{fig_mf}) is determined by the zero crossing of $C_{\rm min}$ at critical $N_{t,c}$. In Fig.\ref{fig_phase}(a), we show $C_{\rm min}$ at $N_{t,c}$ as a function of $g_{12}/g$ with a given $g'/g=-\sqrt{2}/2$. We can see that $C_{\rm min}$ continuously decreases as $g_{12}$ becomes more attractive, and reduces to zero at $g_{12}/g\approx-4$, which separates region II(A) from II(B) on the vertical line in Fig.\ref{fig_mf}. In fact, $C_{\rm min}$ can be estimated from the minimization of $\epsilon_{\rm mf}$, which gives
\begin{equation}
C_{\rm min}^{(0)}=\frac{g+g_{12}-2g'}{g-g'}.   \label{c_eq}
\end{equation}
In Fig.\ref{fig_phase}(a), we can see that Eq.(\ref{c_eq}) well fits $C_{\rm min}$ in region I, but deviates visibly as entering region II(A) when the system  is far away from the mean-field collapse line. 

In Fig.\ref{fig_phase}(b), we further map out the phase diagram highlighting the Borromean droplet(``BD") in the $(g_{12},N)$ plane, taking a fixed $g'/g=-\sqrt{2}/2$. 
To summarize, the Borromean droplet occurs at $N>N_{t,c}$ in region I and $N\in(N_{t,c},N_{b,c})$ in region II(A). We can see that  both $N_{t,c}$ and $N_{b,c}$ decrease as $g_{12}$ gets more attractive.

Now we analytically estimate $N_{t,c}$. First, we investigate the equilibrium density of Borromean droplet by enforcing the zero pressure $P=\sum_i n_i\partial\epsilon/\partial n_i-\epsilon=0$, where $\epsilon=\epsilon_{\rm mf}+\epsilon_{\rm LHY}$. Utilizing $\epsilon_{\rm mf}=gn_1^2f_1$ with $f_1=C^2/2+2Cg'/g +1+g_{12}/g$, and the LHY energy at the mean-field collapse line\cite{supple} $\epsilon_{\rm LHY}=8/(15\pi^2)(gn_1)^{5/2}f_2$, with $f_2=(1+g_{12}/g+C)^{5/2}+(1-g_{12}/g)^{5/2}$, we obtain the density of component-$i$ in the ternary droplet
\begin{equation}
n_{t,i}^{(0)}=\eta_i\frac{25\pi}{1024a^3}\left(\frac{f_1}{f_2}\right)^2. \label{n_eq}
\end{equation}
Here $a=mg/(4\pi)$, $\eta_1=\eta_2=1$ and $\eta_3=C_{\rm min}^{(0)}$. Further, based on (\ref{c_eq},\ref{n_eq}) and the single-mode assumption $\Psi_i({\cp r})=\sqrt{n_{t,i}^{(0)}}\psi({\cp r})$, the coupled GP equations (\ref{GP}) can be reduced to a single one similar to that in the binary case\cite{Petrov}. This results in the following critical number at the transition between ternary droplet and gas phase:
\begin{equation}
N_{t,c}^{(0)}=(2+C)^{5/2}\left(\frac{3}{2}\right)^{3/2}\frac{4\tilde{N}_c}{5\pi^2}\frac{f_2}{f_1^2},\label{N3c}
\end{equation}
with $\tilde{N}_c=18.65$ (at the vanishing of droplet solution). In Fig.\ref{fig_Borro}(c), we show that (\ref{N3c}) fits numerical $N_{t,c}$ qualitatively well over the parameter regime considered. When $C=0$, (\ref{n_eq},\ref{N3c}) recover the results for binary droplets\cite{Petrov}.

{\it Droplet phase separation.} 
Outside the Borromean regime, both the ternary and binary droplets can survive and will directly compete with each other.  Here we will demonstrate an interesting phenomenon of droplet phase separation as below.

\begin{figure}[t]
\includegraphics[width=7cm]{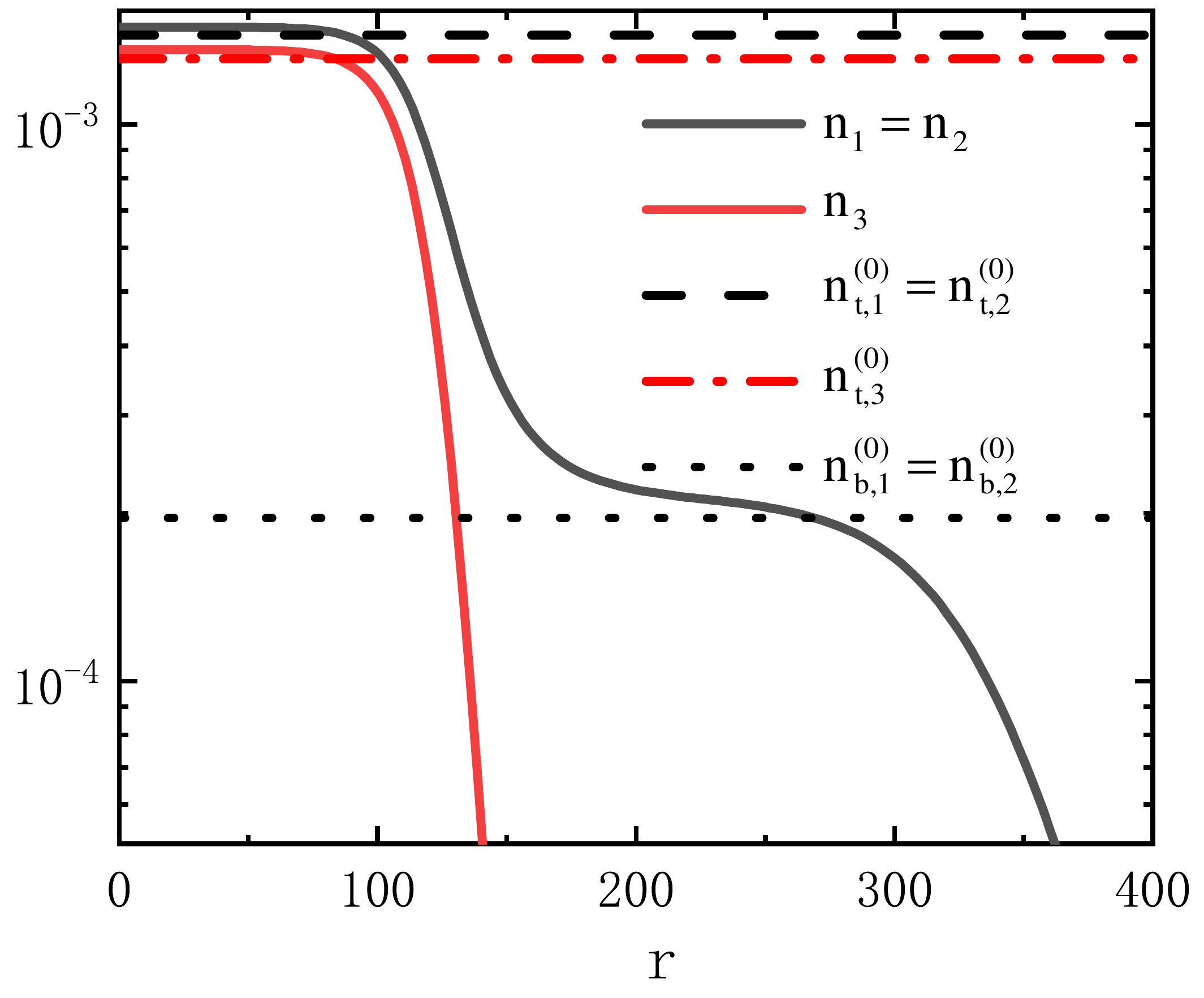}
\caption{(Color online.) Density profile displaying the phase separation between ternary ($1\&2\&3$)  and binary ($1\&2$) droplets. Here $g'/g=-\sqrt{2}/2, g_{12}/g=-1.5, N=10^5, C=0.2$, corresponding to triangular point in Fig.\ref{fig_Borro}(b). 
Horizontal lines show function fits to the equilibrium densities  of ternary (Eq.\ref{n_eq}) and binary (from Ref.\cite{Petrov}) droplets. The length and density units are, respectively, $a$ and $1/a^3$ $(a=mg/(4\pi))$. } \label{fig_PS}
\end{figure}

We consider the region II(A) with a large $N(>N_{b,c})$, i.e., above the ``BD" region in Fig.\ref{fig_phase}(b). In this case, as shown in Fig.\ref{fig_mf}(b), the droplet solution can appear in a reasonably broad range of $C$ and its energy minimum occurs at $C_{\rm min}\neq 0$. Among these $C$-values, $C=0$ and $C=C_{\rm min}$ represent two typical solutions corresponding to, respectively, the binary($1\&2$) and ternary($1\&2\&3$) droplets.  We find that for certain intermediate $C$,  two types of droplet can coexist in the form of phase separation. As shown in Fig.\ref{fig_PS},  it is manifested by two different plateaus in the density profile that well fit the equilibrium densities of ternary and binary droplets. Specifically,  ($1\&2\&3$) droplet occupies the center with $C\sim C_{\rm min}$ and a higher density, while ($1\&2$) droplet stays at edge with $C=0$ and a lower density. Such distribution is believed to lower the surface energy the most.

Here we estimate the $(N,C)$ parameter regime to support the droplet phase separation.  For given $N$ and $C$, we have $N_3=NC/(C+2)$ and $N_1=N_2=N/(C+2)$. Since the full number of component 3, together with part of $1\&2$ components, occupy at the center to form ($1\&2\&3$) droplet with number ratio $C_{\rm min}$, the total number of ternary droplet is $N_{t}=N_3(C_{\min}+2)/C_{\rm min}$. The rest $1\&2$ components are left to form the binary droplet with number $N_{b}=2N_1-2N_3/C_{\rm min}$. The appearance of two types of droplets thus requires $N_{t}>N_{t,c}$ and $N_{b}>N_{b,c}$, setting the constraint for allowed $N$-$C$ values. For any given $N(>N_{b,c})$, the allowed $C$ is within certain window $(C_L,C_H)$, as shown by yellow region in  Fig.\ref{fig_Borro}(b). For very large $N$, we have $C_L\sim 0$ and $C_H\sim C_{\rm min}$, and thus the droplet phase separation can occur for any $C\in(0, C_{\rm min})$.

{\it Summary and discussion.} In summary,  we have shown that the droplet formation in three-component bosons can exhibit much richer physics than that in binary systems, including the enhanced density fluctuations towards mean-field collapse, the occurrence of Borromean droplet, as well as the competition and phase separation between different types of droplets. Though we have focused on the equal mass case with certain coupling symmetries (\ref{simp}), the underlying physics revealed here is robust and can be extended to more general case of mass ratios and coupling strengths.

For the experimental detection of our results, especially regions  I and II(A) in Fig.\ref{fig_mf}, one would need the three-component bosons to hold (i) all repulsive intra-species couplings and (ii) at least two inter-species couplings to be attractive. A good starting point is to first find 
$|1\rangle$ and $|2\rangle$ that may support a binary droplet, such as two $|F=1\rangle$ hyperfine states $|m_F=0\rangle$ and $|m_F=-1\rangle$ of $^{39}$K atom\cite{K39_theory,K39_expt} used to observe quantum droplet near $B_0\sim 57$G\cite{Tarruell_1,Tarruell_2,Inguscio}. Then a third component $|3\rangle$ is required to own a repulsive coupling itself and interact attractively with $|1\rangle$ or $|2\rangle$ near $B_0$.  Key observations would include the emergence of ternary droplet without any binary ones, and the spontaneous phase separation when they coexist. 
As more and more Feshbach resonances are explored between hetero-nuclear bosons, such as $^{41}$K-$^{87}$Rb\cite{K-Rb1}, $^{39}$K-$^{87}$Rb\cite{K-Rb}, $^{23}$Na-$^{87}$Rb\cite{Na-Rb}, $^{39}$K-$^{133}$Cs\cite{K-Cs}, etc, it would be promising in future to find the proper three-component mixtures and detect the droplet physics therein.

Finally, we discuss the possibility of droplet formation with high-order Borromean structure, i.e., following the  {\it Brunnian} ring\cite{brunn1,brunn2,brunn3}. We remark that it is possible to form the 
$n$-th order Brunnian droplet, where only the $n$-component bosons together can form a self-bound state while any $m(<n)$-component cannot. The underlying mechanism resembles that of Borromean droplet revealed in this work, i.e., an extra component brings additional attractive force to the system via density fluctuations. 
Consider a simple case where $g_{ii}=g$ ($i=2,...n$) and all zero inter-species couplings  except $g_{1i}=g'$, the $n$-th order Brunnian droplet will occur if $gg_{11}/g'^2\in (n-2,n-1)$.
This shows the power of collective many-body effect in engineering bound states with a diversely fascinating structure.

\acknowledgments
{\bf Acknowledgment.}
The work is supported by the National Key Research and Development Program of China (2016YFA0300603, 2018YFA0307600), the National Natural Science Foundation of China  (No.12074419), and the Strategic Priority Research Program of Chinese Academy of Sciences (No. XDB33000000).

\clearpage

\begin{widetext}

{\bf Supplemental Materials}

\bigskip

In this Supplemental Material, we provide more details on the derivations of  mean-field instability and the LHY energy for three-component bosons.

\bigskip

{\bf I. Mean-field instability against density fluctuations}

\bigskip

As discussed in the main text, the mean-field stability in the presence of density fluctuations is determined by the second-order variation of mean-field energy $\delta^2 \epsilon_{\rm mf}$, as shown by Eq.(6) in the main text. Further diagonalizing $\delta^2 \epsilon_{\rm mf}$ leads to:
\begin{equation}
\delta^2 \epsilon_{\rm mf}=\tilde{g}_1 \delta \tilde{n}_1^2+\tilde{g}_2 \delta \tilde{n}_2^2+\tilde{g}_3 \delta \tilde{n}_3^2,
\end{equation}
where the eigen-fluctuation-energy reads
\begin{eqnarray}
\tilde{g}_1&=&\frac{g-g_{12}}{2},\\
\tilde{g}_2&=&\frac{g+g_{33}+g_{12}-\Delta}{4},\\
\tilde{g}_3&=&\frac{g+g_{33}+g_{12}+\Delta}{4},
\end{eqnarray}
with $\Delta=\sqrt{8g'^2+(g+g_{12}-g_{33})^2}$. The according eigen-fluctuation density reads
\begin{eqnarray}
{\delta\tilde{ n}}_1&=&\frac{1}{\sqrt{2}} (\delta n_1-\delta n_2),\\
{\delta\tilde{ n}}_2&=&\sqrt{\frac{2g'^2}{\Delta(\Delta+(g+g_{12}-g_{33}))}}\left(\delta n_1+\delta n_2-\frac{\Delta +(g+g_{12}-g_{33})}{2 g'}\delta n_3\right),\\
{\delta\tilde{ n}}_3&=&\sqrt{\frac{2g'^2}{\Delta(\Delta-(g+g_{12}-g_{33}))}}\left(\delta n_1+\delta n_2+\frac{\Delta -(g+g_{12}-g_{33})}{2 g'}\delta n_3\right),
\end{eqnarray}
Here ${\delta\tilde{ n}}_1$ is just $\delta n_-$ in the main text, and ${\delta\tilde{ n}}_2,{\delta\tilde{ n}}_3$ are the two eigen-modes due to coupling between $\delta n_+$ and $\delta n_3$.

\begin{figure}[h]
  \centering
  \includegraphics[width=10cm]{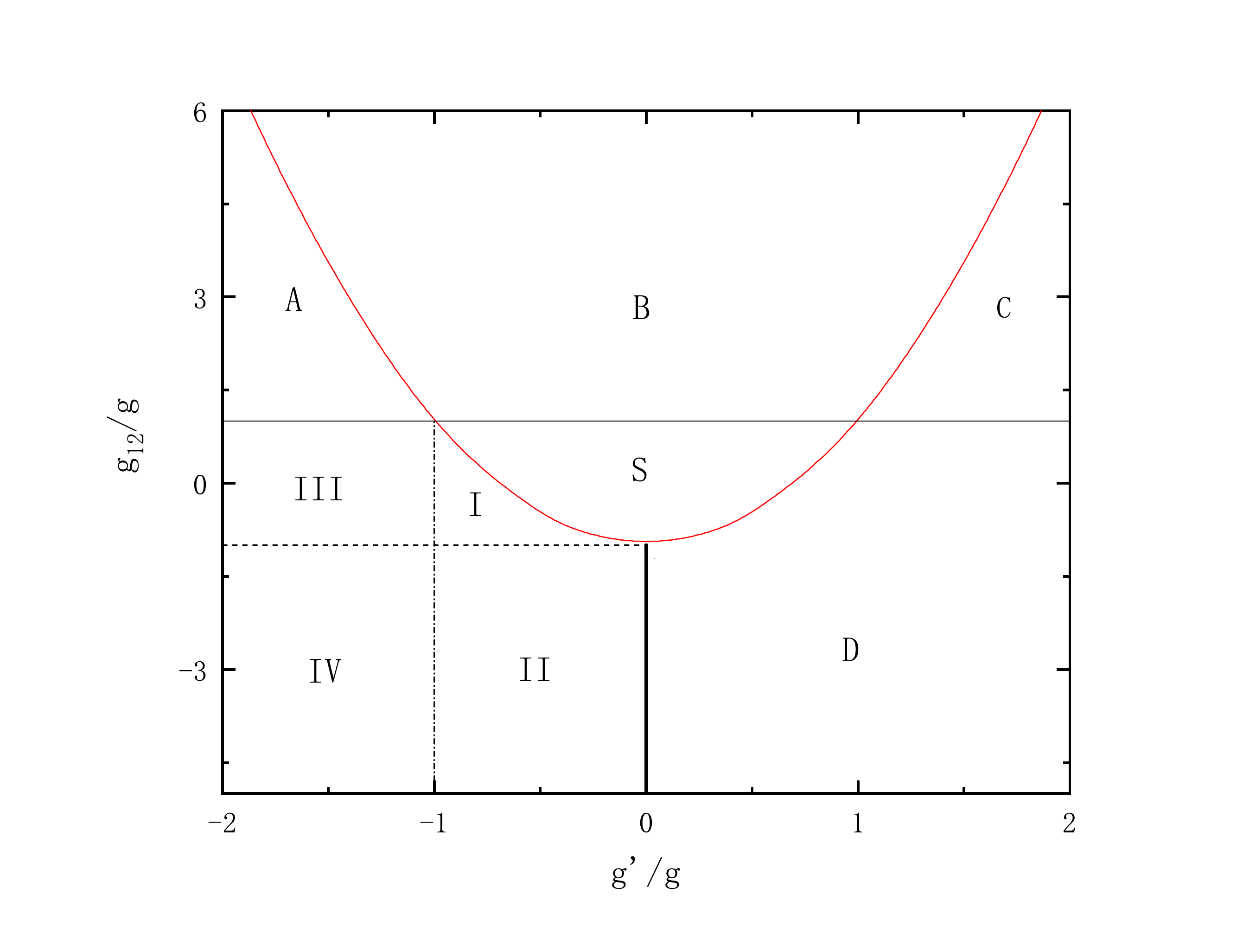}
  \caption{Mean-field phase diagram for three-component bosons with couplings $g_{ii}=g (i=1,2,3),\ g_{13}=g_{23}=g'$. The mean-field stable region is labeled as ``S". The rest ones, labeled as ``I,II,III,IV" and ``A,B,C,D", are all mean-field unstable regions. }\label{fig_SM_mf}
\end{figure}

To ensure the mean-field stability, all the three eigen-modes should have positive energies, i.e., all $\tilde{g}_i>0$. This leads to the stability condition presented as Eq.7 in the main text. In Fig.\ref{fig_SM_mf}, we show the mean-field phase diagram for three-component bosons with couplings $g_{ii}=g (i=1,2,3),\ g_{13}=g_{23}=g'$. The mean-field stable region is labeled as ``S", and the rest unstable regions are labeled as ``I,II,III,IV" and ``A,B,C,D". Next we will analyze the fate of a homogeneous ternary mixture under density fluctuations in these unstable regions.

\begin{itemize}
\item Region A: $\tilde{g}_{1}<0,\tilde{g}_{2}<0,\tilde{g}_{3}>0$. To lower the energy, the system tends to have finite $\delta \tilde{n}_1,\delta \tilde{n}_3$ and zero $\delta \tilde{n}_2$. Thus one has $\delta n_1\neq\delta n_2$ and $\delta n_1+\delta n_2\propto\delta n_3$. This implies that the density fluctuations lead to a collapse between 3 and (1,2) mixture, while (1,2) itself tends to phase separate.
\item Region B: $\tilde{g}_{1}<0,\tilde{g}_{2}>0,\tilde{g}_{3}>0$, which leads to a finite $\delta \tilde{n}_1$ while $\delta \tilde{n}_2=\delta \tilde{n}_3=0$. Hence we have $\delta n_1=-\delta n_2,\delta n_3=0$, which means that  (1,2) phase separate, and 3 remains stable.
\item Region C: $\tilde{g}_{1}<0,\tilde{g}_{2}<0,\tilde{g}_{3}>0$, which leads to finite $\delta \tilde{n}_1,\delta \tilde{n}_2$ while $\delta \tilde{n}_3=0$. This gives $\delta n_1\neq\delta n_2$, $\delta n_1+\delta n_2\propto-\delta n_3$. Therefore there is a phase separation between 3 and (1,2), and 1, 2 themselves also phase separate.
\item Region D: $\tilde{g}_{1}>0,\tilde{g}_{2}<0,\tilde{g}_{3}>0$, which leads to finite $\delta \tilde{n}_2$ and $\delta \tilde{n}_1=\delta \tilde{n}_3=0$. This gives $\delta n_1=\delta n_2$, $\delta n_1+\delta n_2\propto-\delta n_3$, and therefore there is a phase separation between 3 and (1,2), while 1, 2 themselves collapse.
\item Region I, II, III, IV: $\tilde{g}_{1}>0,\tilde{g}_{2}<0,\tilde{g}_{3}>0$, which, similar to region D, leads to finite $\delta \tilde{n}_2$ and $\delta \tilde{n}_1=\delta \tilde{n}_3=0$. However, $g'$ in this region has an opposite sign with region D. Therefore we have $\delta n_1=\delta n_2$ and $\delta n_1+\delta n_2\propto\delta n_3$. This means that 3 and (1,2) tend to collapse, and 1,2 themselves also collapse. These are the regions that the three components all undergo collapse under density fluctuations.
  \end{itemize}

Above results are summarized in Table.\ref{tab1}, where we have listed the parameter regime for each region and the tendency of three components with respect to density fluctuations, including collapse(C), phase separation(PS) and stable(S).

\begin{table*}
  \centering
  \begin{tabular}{c|c|c|c}
     \hline
     \hline
     Region Label &$g_{12}/g$&$g'/g$&3$-$(1,2);1$-$2\\
     \hline
     A & $(1,\infty)$ & $(-\infty,-g'_{c})$&C;PS \\
     B & $(1,\infty)$ & $(-g'_{c},g'_{c})$&S;PS \\
     C & $(1,\infty)$ & $(g'_{c},\infty)$&PS;PS \\
     D & $(-\infty,1)$ &$(g'_{c},\infty)$&PS;C \\
     S &(-1,1)&$(-g'_{c},g'_{c})$&S;S\\
     I & $(-1,1)$ &$(-1,-g'_{c})$&C;C \\
     II &$(-\infty,-1)$ &$(-1,0)$&C;C \\
     III& $(-1,1)$ &$(-\infty,-1)$&C;C \\
     IV &$(-\infty,-1)$ &$(-\infty,-1)$&C;C \\

     \hline
   \end{tabular}
  \caption{Mean-field instability for each region in Fig.\ref{fig_SM_mf}. The last column lists the tendency of the system due to density fluctuations between 3 and (1,2) (``3$-$(1,2)"), and between 1 and 2 (``1$-$2"). ``C", ``PS",``S" respectively stand for collapse, phase separation and stable. Here $g'_c=\sqrt{\frac{1+g_{12}/g}{2}}$.}\label{tab1}
\end{table*}

\bigskip

{\bf II. Lee-Huang-Yang energy for three-component bosons}

\bigskip

Based on the standard Bogoliubov theory, we expand the field operator as:
\begin{equation}
	\begin{split}
		\Psi_i=\sqrt{n_i}+\sum_{{\cp k}\neq 0}\frac{1}{\sqrt{V}}\exp(i{\cp k}{\cp r})\theta_{i{\cp k}}
	\end{split}
\end{equation}
Here $\theta_{i{\cp k}}$ is the fluctuation operator for component-$i$ boson at momentum ${\cp k}$. Then the Hamiltonian can be transformed to the bilinear form $H/V=(1/V)\sum_{\cp k}\phi^\dagger h_{\cp k}\phi+\epsilon_{\rm mf}-(1/V)\sum_{i{\cp k}}(\epsilon_{i{\cp k}}+g_{ii}n_i)$, where $\phi=\left(\theta_{1{\cp k}},\theta_{2{\cp k}},\theta_{3{\cp k}},\theta_{1-{\cp k}}^\dagger,\theta_{2-{\cp k}}^\dagger,\theta_{3-{\cp k}}^\dagger\right)^T$, and

\begin{equation}
	\begin{split}
		h_{\cp k}=\left(
		\begin{array}{cccccc}
			\epsilon_{1{\cp k}}+g_{11}n_1&g_{12}\sqrt{n_1 n_2}&g_{13}\sqrt{n_1 n_3}&g_{11}n_1&g_{12}\sqrt{n_1 n_2}&g_{13}\sqrt{n_1 n_3}\\
			g_{12}\sqrt{n_1n_2}&\epsilon_{2{\cp k}}+g_{22}n_2&g_{23}\sqrt{n_2 n_3}&g_{12}\sqrt{n_1n_2}&g_{22}n_2&g_{23}\sqrt{n_2 n_3}\\
g_{13}\sqrt{n_1n_3}&g_{23}\sqrt{n_2 n_3}&\epsilon_{3{\cp k}}+g_{33}n_3&g_{13}\sqrt{n_1n_3}&g_{23}\sqrt{n_2 n_3}&g_{33}n_3\\			
g_{11}n_1&g_{12}\sqrt{n_1n_2}&g_{13}\sqrt{n_1 n_3}&\epsilon_{1{\cp k}}+g_{11}n_1&g_{12}\sqrt{n_1n_2}&g_{13}\sqrt{n_1 n_3}\\
			g_{12}\sqrt{n_1n_2}&g_{22}n_2&g_{23}\sqrt{n_2 n_3} &g_{12}\sqrt{n_1n_2}&\epsilon_{2{\cp k}}+g_{22}n_2&g_{23}\sqrt{n_2 n_3}\\
g_{13}\sqrt{n_1n_3}&g_{23}\sqrt{n_2 n_3}&g_{33}n_3&g_{13}\sqrt{n_1n_3}&g_{23}\sqrt{n_2 n_3}&\epsilon_{3{\cp k}}+g_{33}n_3\\			
		\end{array}
			\right)
	\end{split}
\end{equation}

The LHY energy can then be derived as Eq.3 in the main text, with $E_{i{\cp k}} \ (i=1,2,3)$ the three Bogoliubov excitation energies. After straightforward algebra, we find that $E_{i{\cp k}}^2$  are the three roots of following equation
\begin{equation}
x^3+bx^2+cx+d=0,
\end{equation}
where
\begin{eqnarray}
b&=&-\sum_i \omega_{i}^2,\\
c&=&\sum_{i<j}\left((\omega_{i}\omega_{j})^2-4g_{ij}^2n_in_j\epsilon_{i{\cp k}}\epsilon_{j{\cp k}}\right),\\
d&=&-(\omega_{1}\omega_{2}\omega_{3})^2-16\epsilon_{1{\cp k}}\epsilon_{2{\cp k}}\epsilon_{3{\cp k}}g_{12}g_{23}g_{13}n_1n_2n_3+\sum_{i<j, l\neq(i,j)}4\epsilon_{i{\cp k}}\epsilon_{j{\cp k}}n_in_jg_{ij}^2\omega_{l}^2.
\end{eqnarray}
Here $\omega_{i}=\sqrt{\epsilon_i^2+2g_{ii}n_i\epsilon_{i{\cp k}}}$ ($i=1,2,3$) are the Bogoliubov spectra for the individual components. Under the equal mass case $m_1=m_2=m_3=m, \epsilon_{\cp k}=\epsilon_{i{\cp k}}$ and coupling symmetry $g_{11}=g_{22}=g$ and $g_{23}=g_{13}=g'$, we can analytically write down the three Bogoliubov energies at the mean-file collapse line $g'=-\sqrt{g_{33}(g+g_{12})/2}$:
\begin{equation}
\begin{split}
E_{1{\cp k}}&=\epsilon_{\cp k}\\
E_{2{\cp k}}&=\left[\epsilon_{\cp k}^2+\epsilon_{\cp k}\left(g(n_1+n_2)+g_{33}n_3-\sqrt{g^2(n_{2}-n_1)^2+(g_{33}n_3+2g_{12}n_1)(g_{33}n_3+2g_{12}n_2)}\right)\right]^{1/2}\\
E_{3{\cp k}}&=\left[\epsilon_{\cp k}^2+\epsilon_{\cp k}\left(g(n_1+n_2)+g_{33}n_3+\sqrt{g^2(n_{2}-n_1)^2+(g_{33}n_3+2g_{12}n_1)(g_{33}n_3+2g_{12}n_2)}\right)\right]^{1/2}
\end{split}
\end{equation}
One can see that at the mean-file collapse line, one mode $E_{1{\cp k}}$ becomes quadratic while the other modes are still linear at low energy. In this case, LHY energy can be simplified as
\begin{equation}
\begin{split}
\varepsilon_{\rm LHY}=&\frac{\sqrt{2}}{15\pi^2}\left[(\alpha+\beta)^{5/2}+(\alpha-\beta)^{5/2}\right]
\end{split}
\end{equation}
where $\alpha=g(n_1+n_2)+g_{33}n_3,\beta=\sqrt{g^2(n_{2}-n_1)^2+(g_{33}n_3+2g_{12}n_1)(g_{33}n_3+2g_{12}n_2)}$. When $g_{33}=g,n_1=n_2$ and $n_3/n_1=C$, the LHY energy is given by $\epsilon_{\rm LHY}=\frac{8}{15\pi^2}f_2$, with $f_2=(1+g_{12}/g+C)^{5/2}+(1-g_{12}/g)^{5/2}$. This is used to estimate $n_i^{(0)}$(Eq.10) in the main text.

For the parameters in regions I and II, the Bogolyubov energies can be complex and thus the LHY energy can be complex. In our numerical calculation, we have simply ignored the imaginary part of LHY energy and only kept its real part.

\end{widetext}


\begin{thebibliography}{99}



\bibitem{halo1} M. V. Zhukov, B. V. Danilin, D. V. Fedorov, J. M. Bang, I. S. Thompson, and J. S. Vaagen, Phys. Rep. {\bf 231}, 151 (1993).

\bibitem{halo2} D. V. Fedorov, A. S. Jensen, and K. Riisager, Phys. Rev. C {\bf 49}, 201 (1994).

\bibitem{Efimov_review1}E. Braaten and H.-W. Hammer, Phys. Rep. {\bf428}, 259 (2006).
\bibitem{Efimov_review2}C.H. Greene, P. Giannakeas, and J. P\'{e}rez-R\'{i}os, Rev. Mod. Phys. {\bf 89}, 035006 (2017).
\bibitem{Efimov_review3}P. Naidon, S. Endo, Rep. Prog. Phys. {\bf 80}, 056001(2017).

\bibitem{Efimov_Exp0} 
T. Kraemer, M. Mark, P. Waldburger, J. G. Danzl, C. Chin, B. Engeser, A. D. Lange, K. Pilch, A. Jaakkola, H.-C. N\"{a}gerl and R. Grimm, Nature {\bf 440}, 315 (2006).

\bibitem{Efimov_Exp1}T. B. Ottenstein, T. Lompe, M. Kohnen, A. N. Wenz,
and S. Jochim, Phys. Rev. Lett. {\bf 101}, 203202 (2008). 

\bibitem{Efimov_Exp2}J. R. Williams, E. L. Hazlett, J. H. Huckans, R. W. Stites, Y. Zhang, and K. M. O'Hara, Phys. Rev. Lett. {\bf 103}, 130404 (2009). 

\bibitem{Efimov_Exp3}
M. Zaccanti, B. Deissler, C. D'Errico, M. Fattori, M. Jona-Lasinio, S. M\"{u}ller, G. Roati, M. Inguscio and G. Modugno, Nat. Phys. {\bf5}, 586 (2009). 

\bibitem{Efimov_Exp4}
N. Gross, Z. Shotan, S. Kokkelmans and L. Khaykovich, Phys. Rev. Lett. {\bf103}, 163202 (2009). 

\bibitem{Efimov_Exp5}
S. E. Plooack, D. Dries and R. G. Hulet, Science {\bf326}, 1683 (2009). 




\bibitem{Efimov_Exp9}
M. Berninger, A. Zenesini, B. Huang, W. Harm, H.-C. N\"{a}gerl, F. Ferlaino, R. Grimm, P. S. Julienne and J. M. Hutson, Phys. Rev. Lett. {\bf107}, 120401 (2011). 

\bibitem{Efimov_Exp10}
R. J. Wild, P. Makotyn, J. M. Pino, E. A. Cornell and D. S. Jin,
Phys. Rev. Lett. {\bf108}, 145305 (2012). 

\bibitem{Richard}J.-M. Richard and S. Fleck, Phys. Rev. Lett. {\bf 73}, 1464 (1994).
\bibitem{Moszkowski}S. Moszkowski, S. Fleck, A. Krikeb, L. Theussl, J.M.
Richard, and K. Varga, Phys. Rev. A {\bf 62}, 032504 (2000)

\bibitem{Nielsen}E. Nielsen, D. V. Fedorov, and A. S. Jensen, Few-Body Systems
{\bf 22}, 15 (1999);
\bibitem{Volosniev}A. G. Volosniev, D. V. Fedorov, A. S. Jensen, and N. T. Zinner, Eur. Phys. J. D {\bf 67}, 95 (2013).
\bibitem{Volosniev2} A. G. Volosniev, D. V. Fedorov, A. S. Jensen, and N. T. Zinner, arxiv: 1312.6535.

\bibitem{Cui}X. Cui, W. Yi, Phys. Rev. X {\bf 4}, 031026 (2014).

\bibitem{helium_expt}S. Grebenev, J. P. Toennies, A. F. Vilesov, Science {\bf 279}, 2083 (1998).

\bibitem{helium_theory}M. Barranco, R. Guardiola, S. Hernandez, R. Mayol, J. Navarro, J. Low Temp. Phys. {\bf 142}, 1(2006)

\bibitem{Huang}K. Huang, Phys. Rev. {\bf 115}, 765 (1959); Phys. Rev. {\bf 119}, 1129 (1960).

\bibitem{Petrov}D.S. Petrov, Phys. Rev. Lett. {\bf 115}, 155302 (2015).


\bibitem{Pfau_1}I. Ferrier-Barbut, H. Kadau, M. Schmitt, M. Wenzel, and T. Pfau, Phys. Rev. Lett. {\bf 116}, 215301 (2016).
\bibitem{Pfau_2}M. Schmitt, M. Wenzel, F. B{\"{o}}ttcher, I. Ferrier-Barbut, and T. Pfau,  Nature {\bf 539}, 259 (2016).
\bibitem{Pfau_3}I. Ferrier-Barbut, M. Schmitt, M. Wenzel, H. Kadau, and T. Pfau, J. Phys. B {\bf 49}, 214004 (2016).
\bibitem{Ferlaino}L. Chomaz, S. Baier, D. Petter, M.J. Mark, F. W{\" a}chtler, L. Santos, and F. Ferlaino, Phys. Rev. X {\bf 6}, 041039 (2016).
\bibitem{Modugno}L. Tanzi, E. Lucioni, F. Fama, J. Catani, A. Fioretti, C. Gabbanini, R. N. Bisset, L. Santos, and G. Modugno, Phys. Rev. Lett. {\bf 122}, 130405 (2019).
\bibitem{Pfau_4}F. B{\" o}ttcher, J.-N. Schmidt, M. Wenzel, J. Hertkorn, M. Guo, T. Langen, and T. Pfau, Phys. Rev. X {\bf 9}, 011051 (2019).
\bibitem{Ferlaino_2}L. Chomaz, D. Petter, P. Ilzh{\"{o}}fer, G. Natale, A. Trautmann, C. Politi, G. Durastante, R.M.W. van Bijnen, A. Patscheider, M. Sohmen, M.J. Mark, and F. Ferlaino, Phys. Rev. X {\bf 9}, 021012 (2019).

\bibitem{Tarruell_1}C.R. Cabrera, L. Tanzi, J. Sanz, B. Naylor, P. Thomas, P. Cheiney, and L. Tarruell, Science {\bf 359}, 301 (2018).
\bibitem{Tarruell_2}P. Cheiney, C. R. Cabrera, J. Sanz, B. Naylor, L. Tanzi, L. Tarruell,  Phys. Rev. Lett. {\bf 120}, 135301 (2018). 
\bibitem{Inguscio}G. Semeghini, G. Ferioli, L. Masi, C. Mazzinghi, L. Wolswijk, F. Minardi, M. Modugno, G. Modugno, M. Inguscio, M. Fattori, Phys. Rev. Lett. {\bf 120}, 235301 (2018).
\bibitem{Modugno_2}C. D'Errico, A. Burchianti, M. Prevedelli, L. Salasnich, F. Ancilotto, M. Modugno, F. Minardi, and C. Fort, Phys. Rev. Research {\bf 1}, 033155 (2019). 


\bibitem{Petrov_2}D. S. Petrov and G. E. Astrakharchik,  Phys. Rev. Lett. {\bf 117}, 100401 (2016).
\bibitem{Santos}D. Edler, C. Mishra, F. W{\" a}chtler, R. Nath, S. Sinha, and L. Santos, Phys. Rev. Lett. {\bf 119}, 050403 (2017). 
\bibitem{Jachymski}K. Jachymski and R. Oldziejewski, Phys. Rev. A {\bf 98}, 043601 (2018).
\bibitem{Zin}P. Zin, M. Pylak, T. Wasak, M. Gajda, and Z. Idziaszek, Phys. Rev. A {\bf 98}, 051603(R) (2018).
\bibitem{Buchler}T. Ilg, J. Kumlin, L. Santos, D. S. Petrov, and H. P. B{\" u}chler, Phys. Rev. A {\bf 98}, 051604(R) (2018).
\bibitem{Cui3}X. Cui, Y. Ma, Phys. Rev. Res. {\bf 3}, L012027 (2021).

\bibitem{Cui2}X. Cui, Phys. Rev. A {\bf 98}, 023630 (2018).
\bibitem{Adhikari}S. Adhikari, Laser Phys. Lett {\bf 15}, 095501 (2018).
\bibitem{Rakshit1}D. Rakshit, T. Karpiuk, M. Brewczyk, and M. Gajda, SciPost Phys. {\bf 6}, 079 (2019). 
\bibitem{Rakshit2}D. Rakshit, T. Karpiuk, P. Zin, M. Brewczyk, M. Lewenstein, and M. Gajda, New J. Phys. {\bf 21}, 073027 (2019). 
\bibitem{Wenzel}M. Wenzel, T. Pfau and I. Ferrier-Barbut, Physica Scripta {\bf 93}, 10 (2018). 
\bibitem{Yi}J.-B. Wang, J.-S. Pan, X. Cui, W. Yi, Chin. Phys. Lett. {\bf 37}, 076701 (2020).

\bibitem{Blakie}Joseph C. Smith, D. Baillie, and P. B. Blakie, Phys. Rev. Lett. {\bf 126}, 025302 (2021).
\bibitem{Santos_2}R. N. Bisset, L. A. Pe$\tilde{n}$a Ardila, and L. Santos, Phys. Rev. Lett. {\bf 126}, 025301 (2021).

\bibitem{book} C. J. Pethick and H. Smith, {\it Bose-Einstein Condensation in Dilute Gases}, Cambridge University Press, 2002.

\bibitem{supple}See supplementary material for the more details on the derivation of mean-field instability and the LHY energy.

\bibitem{K39_theory}M. Lysebo and L. Veseth, Phys. Rev A {\bf 81}, 032702 (2010).

\bibitem{K39_expt}S. Roy {\it et al}, Phys. Rev. Lett. {\bf 111}, 053202 (2013).

\bibitem{K-Rb1}G. Thalhammer, G. Barontini, L. De Sarlo, J. Catani, F. Minardi, and M. Inguscio, Phys. Rev. Lett. {\bf 100}, 210402 (2008).

\bibitem{K-Rb}L. Wacker {\it et al}, Phys. Rev. A {\bf 92}, 053602 (2015).

\bibitem{Na-Rb}F. Wang {\it et al},  J. Phys. B {\bf 49}, 015302 (2015).

\bibitem{K-Cs}M. Gr\"obner {\it et al}, Phys. Rev. A {\bf 95}, 022715 (2017).

\bibitem{brunn1}  H. Brunn, Math. Phys. Klasse, {\bf 22}, 77(1892).    
\bibitem{brunn2} H. Debrunner, Duke Math. J., {\bf 28}, 17 (1961); D.E. Penney, Duke Math. J.,{\bf 36},31(1969); T. Yanagawa, Osaka J. Math., {\bf 1},127 (1964). 
\bibitem{brunn3}Nils A. Baas, D. V. Fedorov, A. S. Jensen, K. Riisager, A. G. Volosniev, N. T. Zinner, Physics of Atomic Nuclei, {\bf 77}, 361 (2014).

\end{thebibliography}
\end{document}